\documentclass[fp,dvipdfmx,twocolumn]{jpsj3}
\usepackage[dvipdfmx]{graphicx}
\usepackage[usenames]{color}
\def\S{Sect. }
\def\deltac{\delta_{\rm c}}
\def\d{{\rm d}}
\def\q{{\rm q}}
\def\msp{m_{\rm sp}}
\def\mspc{m_{\rm sp}^{\rm c}}
\def\v#1{\mib #1}
\def\lambdac{{\lambda_{\rm c}}}
\def\lambdaco{{\lambda_{\rm c0}}}

\def\H{{\mathcal H}}
\newcommand{\aver}[1]{\left\langle {#1} \right\rangle}
\title
{
First Order Transitions Between the Gapped Spin-Liquid and Ferrimagnetic Phases in (1/2,1/2,1) Mixed Diamond Chains with Bond Alternation
}

\author
{
Kazuo Hida\thanks{E-mail address: hida@mail.saitama-u.ac.jp}
}

\inst
{Professor Emeritus, Division of Material Science, Graduate School of Science and Engineering, \\ 
Saitama University, Saitama, Saitama, 338-8570\\ 
}

\recdate
{}

\abst
{
The ground-state phases of mixed diamond chains with bond alternation $\delta$, and ($S, \tau^{(1)}, \tau^{(2)})=(1/2,1/2,1)$, where $S$ is the magnitude of vertex spins, and $\tau^{(1)}$ and $\tau^{(2)}$ are those of apical spins, are investigated. 
 The two apical spins in each unit cell 
are connected 
 by an exchange coupling $\lambda$. The exchange couplings between the apical spins and the vertex spins take the values $1+\delta$ and $1-\delta$ alternatingly. This model has an infinite number of local conservation laws. For large $\lambda$ and $\delta \neq 0$, the ground state is equivalent to that of the spin $1/2$ chain with bond alternation. Hence, the ground state is a gapped spin liquid. This energy gap vanishes for $\delta=0$. With the decrease of  $\lambda$, the ground state undergoes a transition at $\lambda=\lambdaco(\delta)$ to a series of ferrimagnetic phases with a spontaneous magnetization $\msp=1/p$ per unit cell where $p$ is a positive integer. 
 It is found that this transition is a first order transition for $\delta\neq 0$ with a discontinuous change in $\msp$, while no discontinuity is found for $\delta=0$. The critical behaviors of $\msp$ and $\lambdaco(\delta)$ around the critical point $(\delta,\lambda) =(0, \lambdaco(0))$ are also discussed analytically.}
\begin{document}
\maketitle
\section{Introduction}

In low-dimensional frustrated quantum magnets, the interplay of quantum fluctuation and frustration leads to the emergence of various exotic quantum phases.\cite{intfrust,diep} In the one-dimensional cases, the quantized and partial ferrimagnetic phases are often realized in addition to the gapped and gapless spin-liquid phases.

The diamond chain\cite{Takano-K-S,ht2017,kiku2,kiku3,hts2010,hida2019,hida2020,hida2021} is known as one of the simplest examples in which an interplay of quantum fluctuation and frustration leads to a wide variety of ground-state phases. Remarkably, this model has an infinite number of local conservation laws and the ground states can be classified by the corresponding quantum numbers. If the two apical spins have equal magnitudes, the pair of apical spins in each unit cell can form a nonmagnetic singlet dimer and the ground state is a direct product of the cluster ground states separated by singlet dimers.\cite{Takano-K-S,ht2017} Nevertheless, 
 in addition to the spin cluster ground states, various ferrimagnetic states and strongly correlated nonmagnetic states such as the Haldane state are also found when the apical spins form magnetic dimers. In these cases, all the spins collectively form a correlated ground state over the whole chain.  

In the presence of various types of distortion, the spin cluster ground states also turn into highly correlated ground states. Extensive experimental studies have been also carried out on the magnetic properties of the natural mineral azurite that is regarded as an example of distorted spin-1/2 diamond chains.\cite{kiku2,kiku3}

On the other hand, if the magnitudes of the two apical spins are unequal, they cannot form a singlet dimer. Hence, all spins in the chain inevitably form a many-body correlated state. 
 As a simple example of such cases, we investigated the mixed diamond chain with apical spins of magnitude 1 and 1/2, and vertex spins, 1/2 in Ref. \citen{hida2021}. In addition to the nonmagnetic gapless spin liquid phase and the ferrimagnetic phase expected from the Lieb-Mattis theorem,\cite{Lieb-Mattis} we found an infinite series of ferrimagnetic phases with spontaneous magnetizations $\msp=1/p$ where $p$ is a positive integer ($1\leq p < \infty$). The ferrimagnetic phases for $p \geq 2$ are accompanied by the spontaneous translational symmetry breakdown with spatial periodicities of $p$ unit cells. The width and spontaneous magnetization of each ferrimagnetic phase tend to infinitesimal as $\lambda$ tends to the ferrimagnetic-nonmagnetic transition point. Considering the infinitesimal energy scale around this transition point, these series of ferrimagnetic ground states are expected to be fragile against various perturbations such as lattice distortions, randomness, and finite temperature effect. 

In the present work, we investigate the effect of bond alternation $\delta$ on this model. This type of lattice distortion preserves the infinite number of conservation laws in the undistorted diamond chain and a similar series of ferrimagnetic phases are found.  For finite $\delta$, however, it is verified that the maximal value of $p$ is finite and the spontaneous magnetization $\msp$ has a finite discontinuity at the transition point.

This paper is organized as follows. 
 {In \S 2}, the model Hamiltonian is presented. 
 {In \S 3}, the numerical results for the spontaneous magnetization and the first-order transition points are presented. The critical behavior of spontaneous magnetization and the first-order transition point for small $\delta$ is discussed analytically in \S 4.
The last section is devoted to a summary and discussion.

\section{Hamiltonian}

\begin{figure}[t] 
\centerline{\includegraphics[width=6cm]{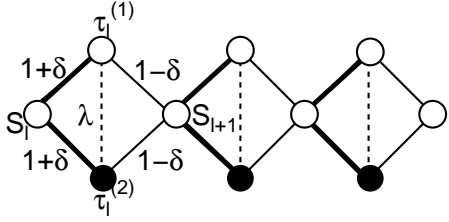}}
\caption{Structure of the diamond chain with bond alternation $\delta$. The spin magnitudes are $S=\tau^{(1)}=1/2$ and $\tau^{(2)}=1$.}
\label{lattice}
\end{figure}

We consider the Hamiltonian 
\begin{align}
{\mathcal H} = &\sum_{l=1}^{L} \Big[(1+\delta)\v{S}_{l}(\v{\tau}^{(1)}_{l}+\v{\tau}^{(2)}_{l}) 
\nonumber\\
&+(1-\delta)(\v{\tau}^{(1)}_{l}+\v{\tau}^{(2)}_{l})\v{S}_{l+1}
+  \lambda\v{\tau}^{(1)}_{l}\v{\tau}^{(2)}_{l}\Big], 
\label{hama}
\end{align}
where $\v{S}_{l}, \v{\tau}^{(1)}_{l}$ and $\v{\tau}^{(2)}_{l}$ are spin operators with magnitudes ${S}_{l}={\tau}^{(1)}_{l}=1/2$ and ${\tau}^{(2)}_{l}=1$. The number of unit cells is denoted by $L$, and the total number of sites is $3L$ if the periodic boundary condition $\v{S}_{L+1}\equiv \v{S}_1$ is employed. Here, the parameters $\lambda$ and $\delta$ control the frustration and bond alternation, respectively, as depicted in Fig. \ref{lattice}. 

The Hamiltonian (\ref{hama}) has a series of local conservation laws. 
To see it, we rewrite Eq. (\ref{hama}) 
 in the form, 
\begin{align}
\H &= \sum_{l=1}^{L} \Big[(1+\delta)\v{S}_{l}\v{T}_{l}+(1-\delta)\v{T}_{l}\v{S}_{l+1}\nonumber\\
&+ \frac{\lambda}{2}\left(\v{T}^2_{l}-\frac{11}{4}\right)\Big],
\label{ham2} 
\end{align}
where the composite spin operators $\v{T}_l$ are defined as 
\begin{align}
\v{T}_{l} \equiv \v{\tau}^{(1)}_{l}+\v{\tau}^{(2)}_{l} 
\quad (l = 1, 2, \cdots ,L). 
\end{align}
Then, it is evident that 
\begin{align}
[\v{T}_l^2, {\mathcal H}] = 0 \quad (l = 1, 2, \cdots , L). \label{eq:cons}
\end{align}
Thus, we have $L$ conserved quantities $\v{T}_l^2$ for all $l$. 
By defining the magnitude $T_l$ of the composite spin $\v{T}_l$ by $\v{T}_l^2 = T_l (T_l + 1)$, we have a 
 set of good quantum numbers $\{T_l; l=1,2,...L\}$ where $T_l=$ 1/2 and 3/2. 
The total Hilbert space of the Hamiltonian (\ref{ham2}) consists of 
separated subspaces, each of which is specified by 
a definite set of $\{T_l\}$, i.e., a sequence of 1/2 and 3/2. 
 A pair of apical spins with $T_l=1/2$ is called a doublet (hereafter abbreviated as d) and that with $T_l=3/2$ a quartet (abbreviated as q). 

\section{Ground-State Phases}

\begin{figure} 
\centerline{\includegraphics[width=7cm]{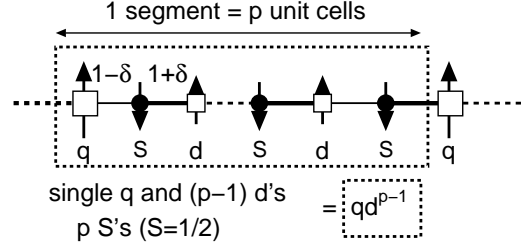}}
\caption{$(\q\d^{p-1})^{\infty}$ configuration.}
\label{fig:periodic_phase}
\end{figure}

\begin{figure} 
\centerline{\includegraphics[width=7cm]{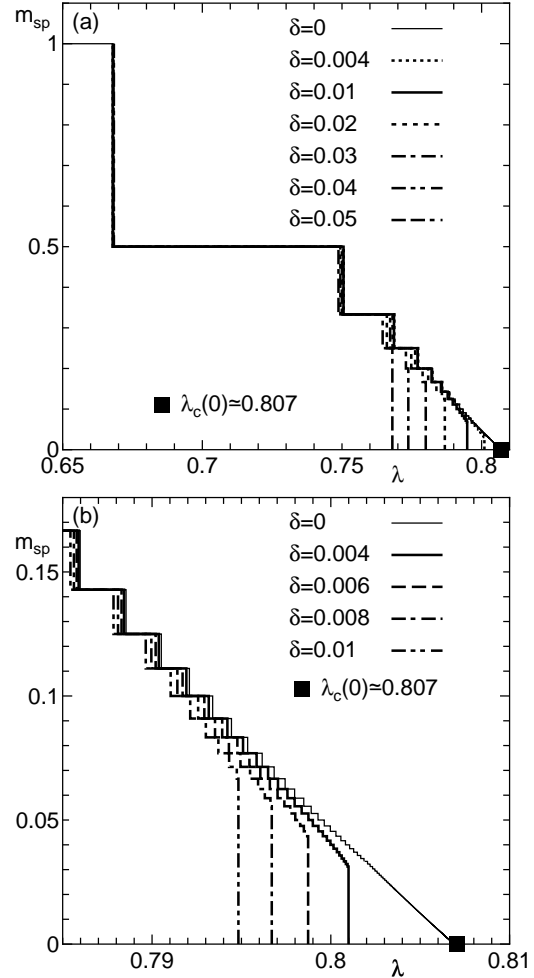}}

\caption{(a) $\lambda$-dependence of $\msp$ calculated by the iDMRG method  for $\delta=0, 0.004, 0.01, 0.02, 0.03, 0.04$, and 0.05. (b) Magnified figure around $\lambda \sim \lambdaco(0)$ for $\delta=0, 0.004, 0.006, 0.008$, and 0.01. The thin solid line for $\delta=0$ is taken from Ref. \citen{hida2021}.}
\label{fig:mag_infinite}
\end{figure}
\begin{figure} 
\centerline{\includegraphics[width=7cm]{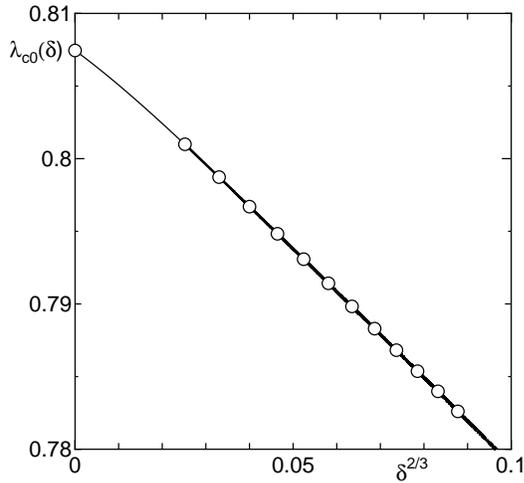}}
\caption{$\delta$-dependence of $\lambdaco(\delta)$ plotted against $\delta^{2/3}$. 
The values obtained from the numerical calculations are indicated by open circles. The thick solid line is obtained from the ground-state energies interpolated from the numerical results. 
 The thin solid line for small $\delta$ is the extrapolation from numerical values of $\lambdaco(\delta)$ for larger $\delta$.}
\label{fig:lambdac}
\end{figure}
\begin{figure} 
\centerline{\includegraphics[width=7cm]{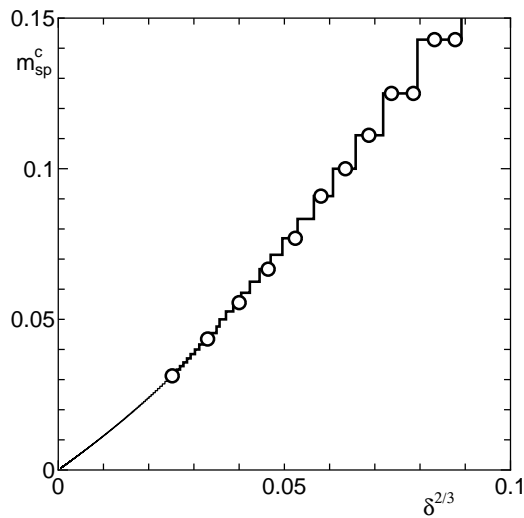}}
\caption{$\delta$-dependence of $\mspc$. The values obtained from the numerical calculations are indicated by open circles.  The thick solid lines are obtained from the ground-state energies interpolated from the numerical results.  The thin solid line for small $\delta$ is drawn by extrapolating the numerical values of the upper and lower ends of the steps from larger $\delta$.
}
\label{fig:mspjump}
\end{figure}

\subsection{Ground states for $\lambda \gg 1$}\label{sect:nonmag}

For $\lambda \gg 1$, $\forall l \ T_l=1/2$. Hence, this model is equivalent to the spin-1/2 antiferromagnetic Heisenberg chain with bond alternation $\delta$ whose ground state is a gapped spin liquid for $\delta \neq 0$. For $\delta=0$, it is a gapless spin liquid.
\subsection{Ground states for $\lambda \ll 1$}\label{sect:full}

For $\lambda \ll 1$, $\forall l \ T_l=3/2$. Hence, this model is equivalent to the spin-1/2-3/2 alternating antiferromagnetic Heisenberg chain whose ground state is a ferrimagnetic state with spontaneous magnetization $\msp=1$ per unit cell according to the Lieb-Mattis theorem.\cite{Lieb-Mattis} Here, $\msp$ is defined by
\begin{align}
\msp=\frac{1}{L}\sum_{l=1}^L(\aver{S^z_l}+\aver{T^z_l}),
\end{align}
where $\aver{}$ denotes the expectation value in the ground state with an infinitesimal symmetry breaking magnetic field in $z$-direction. 
\subsection{Intermediate $\lambda$}

We investigate this regime numerically using the infinite-size DMRG (iDMRG) method. 
The number of states $\chi$ kept in each subsystem in the iDMRG calculation ranged from 240 to 360. We calculate the ground-state energies per unit cell for different configurations of $\{T_l\}$ at the infinite-size fixed point and compare them to find the ground-state configuration. Although we start the iteration with the open boundary condition setting $\v{S}_{L+1}=0$, the boundary condition is irrelevant since we measure the ground-state energies per unit cell for the middle segment of the chain at the infinite-size fixed point.
 However, it is not possible to carry out the calculation for all possible configurations of $\{T_l\}$ for infinite chains.  As plausible candidates of the ground states, we consider the configurations $(\q\d^{p-1})^{\infty}$ with $\msp=1/p$ where $p$ takes positive integer values. The configuration $(\q\d^{p-1})^{\infty}$ consists of an infinite array of segments $\q\d^{p-1}$ with length of $p$ unit cells as depicted in Fig. \ref{fig:periodic_phase}. This phase is called the $(\q\d^{p-1})^{\infty}$ phase. As for the notations of the configurations and phases, we follow those of Ref. \citen{hida2021}. These states are the ground states for $\delta=0$\cite{hida2021}, and we assume that other types of ground states do not emerge by the introduction of a bond alternation. 

The $\lambda$-dependence of $\msp$ is shown in Fig. \ref{fig:mag_infinite} for various values of $\delta$. 
For $\delta\neq 0$, the spontaneous magnetization $\msp$ jumps from $\msp=0$ to a finite value $\mspc(\delta)$ at the nonmagnetic-ferrimagnetic transition point $\lambda=\lambdaco(\delta)$. For $\delta=0$, $\msp$ rises starting from infinitesimal value corresponding to $p \rightarrow \infty$ at the critical value of $\lambda$ given by\cite{hida2021}
\begin{align}
\lambdaco(0)\equiv\lim_{p\rightarrow \infty}\lambdac(p+1,p,\delta=0)\simeq 0.807,\label{eq:lamcdef}
\end{align}
where the boundary between the  $(\q\d^{p-1})^{\infty}$ phase with $\msp=1/p$ and $(\q\d^{p})^{\infty}$ phase with $\msp=1/(p+1)$ is denoted by $\lambdac(p+1,p,\delta)$. The transition point $\lambdaco(\delta)$ shifts to lower values with the increase of $\delta$. This is plotted against $\delta^{2/3}$ in Fig. \ref{fig:lambdac} suggesting the asymptotic behavior  $\lambdaco(\delta)-\lambdaco(0)\propto \delta^{2/3}$. The value of $\lambdaco(0)$ estimated from the extrapolation of $\lambdaco(\delta)$ to $\delta\rightarrow 0$ is slightly larger than the value (\ref{eq:lamcdef}). However, the deviation is within the last digit of 0.001 and would be attributed to the ambiguity in the extrapolation procedure. The spontaneous magnetization $\mspc$  just below the transition point $\lambda=\lambdaco(\delta)$ is also plotted in Fig. \ref{fig:mspjump} against  $\delta^{2/3}$ suggesting the asymptotic behavior  $\mspc \propto \delta^{2/3}$. The ground-state energies per unit cell are numerically calculated by the iDMRG method for each configuration of $\{T_l\}$ and the values of $\delta$ indicated by open circles in Figs. \ref{fig:lambdac} and \ref{fig:mspjump}. The ground-state energies for other values of $\delta$ are estimated by the interpolation from the numerical results to obtain the thick solid lines in Figs. \ref{fig:lambdac} and \ref{fig:mspjump}. The thin solid line in Fig. \ref{fig:lambdac} is calculated by extrapolating the numerical values for larger $\delta$. The thin solid line in Fig.  \ref{fig:mspjump} is drawn by extrapolating the numerical values of the upper and lower ends of the steps from larger $\delta$. .
 These asymptotic behaviors are explained analytically in the next section.

\section{Analytical approach}
 
To examine the critical behavior for $\delta \sim 0$ and $\lambda \sim \lambdaco(0)$ analytically, we start with asymptotic behaviors of the ground-state energies of nonmagnetic and ferrimagnetic phases for small $\delta$.

\subsection{Ground-state energy of the nonmagnetic phase with bond alternation}

For small $\delta$, the ground-state energy of the nonmagnetic phase $E_{\rm N}(\delta, L)$ can be obtained from the well-known result for the spin-1/2 Heisenberg chain with bond alternation $\delta$ as
\begin{align}
E_{\rm N}(\delta, L)&\simeq -\lambda L+ L(2\epsilon_0-C_0\delta^{4/3}),\label{eq:eg_nonmag}
\end{align}
where $\epsilon_0$ and $C_0$ are constants. The energy $\epsilon_0$ corresponds to the ground-state energy of the spin-1/2 antiferromagnetic Heisenberg chain per site. The first term is the contribution from the energy of the apical spins in the d state. The last term is the correction resulting from the bond alternation $\delta$. The $\delta$-dependence of this term is also well-established\cite{cross-fisher,affleck1989,kadanoff1980,giamarchi} on the basis of the SU(2) invariant conformal field theory, although the logarithmic correction is ignored in the present analysis.

\subsection{Ground-state energy of the ferrimagnetic phase} 

In this case, the point $\delta=0$ is not a critical point. Hence, we can expand the ground-state energy with respect to $\delta$. Considering that the energy is an even function of $\delta$, the lowest order correction in $\delta$ is of the order of $\delta^2$ that is smaller than the corresponding correction $\propto \delta^{4/3}$ in the nonmagnetic phase for small $\delta$. Hence, we neglect the effect of bond alternation in the ferrimagnetic phase. 
The ground-state energy $E_{\rm F}(p,\delta,L)$ in the ferrimagnetic $(qd^{p-1})^{\infty}$ phase can be expressed as
\begin{align}
E_{\rm F}(p,\delta,L)&= \frac{1}{2}\lambda\cdot\frac{L}{p}-\lambda \frac{L}{p}(p-1)+ L\epsilon(p),
\end{align}
where the first term is the contribution from the energy of the $L/p$ pairs of apical spins in the q state and the second term is that from the remaining pairs of apical spins in the d state. The energy $\epsilon(p)$ is the ground-state energy per unit cell of the Heisenberg model 
\begin{align}
\H_0 &= \sum_{l=1}^{L} \left[\v{S}_{l}\v{T}_{l}+\v{T}_{l}\v{S}_{l+1}\right],
\label{ham0} 
\end{align}
with the configuration $(qd^{p-1})^{\infty}$. For large $p$, we expand $\epsilon(p)$ up to the second order in $1/p$ as
\begin{align}
\epsilon(p)&\simeq 2\epsilon_0+\frac{C_1}{p}+\frac{C_2}{p^2},
\end{align}
where $C_1$ and $C_2$ are constants.  It should be noted that the limit $p\rightarrow \infty$ corresponds to the uniform spin-1/2 antiferromagnetic Heisenberg chain. Hence, $\epsilon(p)$ should tend to $2\epsilon_0$ in this limit.

\subsection{Phase transition points}

 The transition point $\lambdac(p+1,p,\delta)$ between the  $(\q\d^{p-1})^{\infty}$ and  $(\q\d^{p})^{\infty}$ phases is given by equating the ground-state energies of both phases as
\begin{align}
E_{\rm F}(p,\delta,L)&=E_{\rm F}(p+1,\delta,L),
\end{align}
which yields 
\begin{align}
\lambdac(p+1,p,\delta)&=-\frac{2}{3}\left(\frac{2p+1}{p(p+1)}C_2+C_1\right).
\label{eq:e_ferri0}
\end{align}
Comparing (\ref{eq:e_ferri0}) with the definition of $\lambdaco(0)$ (\ref{eq:lamcdef}), we find
\begin{align}
\lambdaco(0)&=-\frac{2}{3}C_1.
\end{align}
Hence, we have
\begin{align}
E_{\rm F}(p,\delta,L)&=\frac{L}{p}\Big[\frac{3}{2}(\lambda-\lambdaco(0))+\frac{C_2}{p}\Big]\nonumber\\
&+2L\epsilon_0,
\label{eq:e_ferri}
\end{align}
and 
\begin{align}
\lambdac(p+1,p,\delta)-\lambdaco(0)&=-\frac{2}{3}\frac{2p+1}{p(p+1)}C_2.
\label{eq:e_pp}
\end{align}

For finite $\delta$, we may assume the transition to the nonmagnetic phase takes place directly from the $(\q\d^{p-1})^{\infty}$ phase with appropriate $p$. The transition point $\lambdaco(\delta)$ is determined by equating the ground-state energies Eq. (\ref{eq:eg_nonmag}) and Eq. (\ref{eq:e_ferri}) as 
\begin{align}
&\lambdaco(\delta)-\lambdaco (\delta=0)\nonumber\\
&=-\frac{2}{3}\left(pC_0\delta^{4/3}+\frac{C_2}{p}\right),\label{eq:lambdac}
\end{align}
where the spatial periodicity $p$ must satisfy
\begin{align}
\lambdac(p+1,p,\delta)&\geq\lambdaco(\delta)\geq\lambdac(p,p-1,\delta).\label{eq:ineq}
\end{align}
Substituting Eqs.(\ref{eq:e_pp}) and (\ref{eq:lambdac}) into Eq. (\ref{eq:ineq}), we find
\begin{align}
\frac{1}{p(p+1)} &\leq \frac{C_0}{C_2}\delta^{4/3}\leq \frac{1}{p(p-1)}. 
\end{align}
For large enough $p$, this yields
\begin{align}
\frac{1}{p} \simeq \sqrt{\frac{C_0}{C_2}}\delta^{2/3}. \label{eq:pinv}
\end{align}
This implies that the spontaneous magnetization $\msp$ jumps from 0 to $\mspc$ given by
\begin{align}
\mspc&=\frac{1}{p} \simeq \sqrt{\frac{C_0}{C_2}}\delta^{2/3} \propto \delta^{2/3}, \label{eq:msp_ana}
\end{align}
at the transition point. Further, substituting (\ref{eq:pinv}) into (\ref{eq:lambdac}), we find 
\begin{align}
\lambdaco(\delta)-\lambdaco(0)&\simeq 
-\frac{4}{3}\sqrt{C_0C_2}\delta^{2/3}\nonumber\\
  &\propto \delta^{2/3}.\label{eq:lambdac_ana}
\end{align}
These results are consistent with the plots of Fig. \ref{fig:lambdac} and Fig. \ref{fig:mspjump}.

\section{Summary and Discussion}

 The ground-state phases of diamond chains (\ref{hama}) with $(S,\tau^{(1)},\tau^{(2)})=(1/2,1/2,1)$ and bond alternation $\delta$ are investigated. For $\delta \neq 0$, the transition between the gapped spin-liquid and the ferrimagnetic phases is of the first order with discontinuity in spontaneous magnetization. This is in contrast to the case of $\delta=0$, in which $\msp$ rises from the gapless spin liquid phase with an infinitesimal step resulting in the infinite series of ferrimagnetic phases.
The $\delta$-dependence of the transition point $\lambdaco(\delta)$ and that of the spontaneous magnetization $\msp$ at the transition point are examined numerically and analytically.

Our results show that the nature of the ferrimagnetic phase is closely related to that of the neighboring nonmagnetic phase. The infinite series of ferrimagnetic phases  present for $\delta=0$ is truncated at finite $p$ as soon as the spin gap opens in the nonmagnetic side owing to nonvanishing $\delta$. This implies that the infinite series  of ferrimagnetic phases and accompanying infinitesimal magnetization steps are the consequences of the critical nature of the ground state of the uniform spin-1/2 Heisenberg chain as suggested in Ref. \citen{hida2021}. 

In the spin-1  diamond chain with bond alternation $\delta$, the nonmagnetic phase is equivalent to the ground state of the spin-1 Heisenberg chain with bond alternation $\delta$.\cite{hida2020} In this model, an intermediate ferrimagnetic phase is observed in the close neighborhood of the point  $(\lambda,\delta)=(\lambdac(S=1),\deltac(S=1))\simeq (1.0832,0.2598)$ that  corresponds to the endpoint of the Haldane-dimer critical line.\cite{Kato-Tanaka1994,Yamamoto1994,Totsuka1995,Kitazawa-Nomura1997}.  In Ref. \citen{hida2021}, it has been speculated that for $\delta=\deltac(S=1)$ an infinite series of quantized ferrimagnetic phases similar to those discussed in the present model with $\delta=0$ is realized.\cite{hida2021} The behavior for $\delta \simeq \deltac(S=1)$ would be also similar to the present model although the detailed numerical confirmation is too demanding due to the smallness of the width of this region.

As discussed in the Introduction, the series of ferrimagnetic phases predicted for $\delta=0$ are expected to be fragile against various perturbations near $\lambda=\lambdaco(0)$. This implies that a wide variety of exotic phases and phase transitions can emerge from these states in the presence of the perturbations that are expected if materials close to the present model are synthesized experimentally. In this context, it would be important to investigate the effect of lattice distortions that do not preserve the conservation laws (\ref{eq:cons}).\cite{hts2010,hida2019} Also, the effect of randomness would be one of the relevant issues. These studies are left for future investigation.

\acknowledgments

A part of the numerical computation in this work has been carried out using the facilities of the Supercomputer Center, Institute for Solid State Physics, University of Tokyo, and Yukawa Institute Computer Facility at Kyoto University.

\end{document}